\documentclass[]{spie}  
\usepackage[]{graphicx}
\usepackage{epstopdf}
\usepackage{aas_macros}
\usepackage{textcomp}

\title{The next-generation BLASTPol experiment} 

\author{Bradley J. Dober\supit{\textasteriskcentered}\supit{a}, Peter A. R. Ade\supit{b}, Peter Ashton\supit{c}, Francesco E. Angil{\`e}\supit{a}, James A. Beall\supit{d}, Dan Becker\supit{d}, Kristi J. Bradford\supit{e}, George Che\supit{e}, Hsiao-Mei Cho\supit{d}, Mark J. Devlin\supit{a}, Laura M. Fissel\supit{c}, Yasuo Fukui\supit{f}, Nicholas Galitzki\supit{a}, Jiansong Gao\supit{d}, Christopher E. Groppi\supit{e}, Seth Hillbrand\supit{g}, Gene C. Hilton\supit{d}, Johannes Hubmayr\supit{d}, Kent D. Irwin\supit{h}, Jeffrey Klein\supit{a}, Jeff Van Lanen\supit{d}, Dale Li\supit{d}, Zhi-Yun Li\supit{i}, Nathan P. Lourie\supit{a}, Hamdi Mani\supit{e}, Peter G. Martin\supit{j}, Philip Mauskopf\supit{e}, Fumitaka Nakamura\supit{k}, Giles Novak\supit{c}, David P. Pappas\supit{d}, Enzo Pascale\supit{b}, Fabio P. Santos\supit{c}, Giorgio Savini\supit{l}, Douglas Scott\supit{m}, Sara Stanchfield\supit{a}, Joel N. Ullom\supit{d}, Matthew Underhill\supit{e}, Michael R. Vissers\supit{d}, Derek Ward-Thompson\supit{n}
\skiplinehalf
\supit{a}Department of Physics and Astronomy University of Pennsylvania, 209 South 33rd Street, Philadelphia, PA 19104; \\
\supit{b}Department of Physics $\&$ Astronomy, Cardiff University, 5 The Parade, Cardiff CF24 3A A, UK;\\
\supit{c}Department of Physics $\&$ Astronomy, Northwestern University, 2145 Sheridan Road, Evanston, IL 60208; \\
\supit{d}Quantum Electronics and Photonics Division, National Institute of Standards and Technology, 325 Broadway Street, Boulder, CO 80305; \\
\supit{e}Department of Physics, Arizona State University, P.O. Box 871504, Tempe, AZ 85287; \\
\supit{f}Department of Physics and Astrophysics, Nagoya University, Chikusa-ku Nagoya 464-8602, Japan; \\
\supit{g}Department of Physics and Astronomy, California State University, Sacramento, 6000 J Street, Sacramento, CA 95819; \\
\supit{h}Department of Physics, Stanford University, 382 Via Pueblo Mall, Stanford, CA 94305; \\
\supit{i}Department of Astronomy, University of Virginia, P. O. Box 400325, Charlottesville, VA 22904; \\
\supit{j}Department of Astronomy $\&$ Astrophysics, University of Toronto, 50 St. George Street, Toronto, ON M5S 3H4, Canada; \\
\supit{k}National Astronomical Observatory, Mitaka, Tokyo 181-8588, Japan; \\
\supit{l}Department of Physics and Astronomy, University College London, Gower Street, London, WC1E 6BT, UK; \\
\supit{m}Department of Physics and Astronomy, 6224 Agricultural Road, University of British Columbia, Vancouver, BC, V6T 1Z1 Canada; \\
\supit{n}Jeremiah Horrocks Institute of Maths, Physics and Astronomy, University of Central Lancashire, Preston, PR1 2HE, UK;
}

\authorinfo{\supit{{\textasteriskcentered}}Corresponding Author. E-mail: dober@sas.upenn.edu, Telephone: 215-746-3792 \\ Copyright 2014 Society of Photo-Optical Instrumentation Engineers. One
print or electronic copy may be made for personal use only. Systematic
reproduction and distribution, duplication of any material in this paper
for a fee or for commercial purposes, or modification of the content of
the paper are prohibited.}

\begin{document} 
  \maketitle 

\begin{abstract}
 The Balloon-borne Large Aperture Submillimeter Telescope for Polarimetry (BLASTPol) is a suborbital mapping experiment designed to study the role magnetic fields play in star formation. BLASTPol has had two science flights from McMurdo Station, Antarctica in 2010 and 2012. These flights have produced thousands of polarization vectors at 250, 350 and 500 microns in several molecular cloud targets. We present the design, specifications, and progress towards the next-generation BLASTPol experiment (BLAST-TNG). BLAST-TNG will fly a 40$\%$ larger diameter primary mirror, with almost 8 times the number of polarization-sensitive detectors resulting in a factor of 16 increase in mapping speed. With a spatial resolution of $22''$ and four times the field of view (340~arcmin$^{2}$) of BLASTPol, BLAST-TNG will bridge the angular scales between Planck's all-sky maps with $5'$ resolution and ALMA's ultra-high resolution narrow ($\sim 20''$) fields. The new receiver has a larger cryogenics volume, allowing for a 28 day hold time. BLAST-TNG employs three arrays of Microwave Kinetic Inductance Detectors (MKIDs) with 30$\%$ fractional bandwidth at 250, 350 and 500 microns. 
 In this paper, we will present the new BLAST-TNG instrument and science objectives.
\end{abstract}



\section{INTRODUCTION}
\label{sec:intro}  
The BLASTPol experiment was a stratospheric 1.8~m telescope which observed the sky with bolometric detectors operating in three $30\%$ wide bands at 250, 350 and 500~$\mu$m. Polarization sensitivity was achieved via the insertion of polarized grids in front of the detector arrays. Polarization modulation was achieved via a stepped achromatic half-wave plate (HWP). BLASTPol had two Long Duration Balloon (LDB) flights from McMurdo, Antarctica in 2010 and 2012. Some results from these flights are documented elsewhere\cite{matthews2013,poidevin2014,galitzki2014}.

The next generation BLASTPol experiment will improve upon the previous BLASTPol measurements. With a larger 2.5~m primary mirror and almost 8 times more detectors, it will have an increased resolution ($22''$ at $250~\mu$m), four times the field of view (340~arcmin$^{2}$), and up to 16 times the mapping speed of BLASTPol. This performance enhancement will enable us to perform the most detailed study of the role of magnetic fields in the regulation of star formation than ever achieved before.

\section{Scientific Motivation} 
After decades of research, the physical processes regulating star formation remain poorly understood. Large-scale observations of star forming regions provide counts of the number of dense clouds, each of which will eventually evolve into many thousands of stars. However, when simple models of gravitational collapse are applied to the clouds they yield a Galactic star formation rate (SFR) that is many times larger than what is actually observed. Some process or combination of processes must be slowing the collapse of the clouds. The two prevailing theories involve turbulence, which prevents the effective dissipation of energy, and Galactic magnetic fields, which are captured and squeezed by the collapsing cloud providing a mechanism for mechanical support.

In many numerical simulations, magnetic fields dramatically affect both the star formation efficiency and lifetime of molecular clouds\cite{Hennebelle2011,LiNak2004,Li2010}. However observationally, the strength and morphology of magnetic fields in molecular clouds remain poorly constrained. A useful tracer of magnetic fields in star forming regions is polarization. By mapping polarization from dust grains aligned with respect to their local magnetic field, the field orientation (projected on the sky) can be traced. Molecular clouds typically have temperatures of several tens of Kelvin with emission peaking in the submillimeter. Most previous submillimeter polarimetry observations have generally been restricted to bright targets ($>1$~Jy) and small map sizes ($<0.05$~deg$^{2}$)\cite{Dotson2010,Matthews2009}.

One of BLAST-TNG's main goals is to map the magnetic field morphology of a large sample of molecular clouds in order to account for projection effects and obtain robust statistics on magnetic field morphology. A large sample of molecular clouds will allow us to constrain star formation models. BLAST-TNG can provide valuable data that will aid in two major areas of star formation models: how ordered are the magnetic fields in the molecular cloud, and how does that field direction correlate with the filamentary structure within the clouds? Numerical simulations have shown that molecular clouds with strong magnetic fields show less polarization angle dispersion than clouds with weak fields\cite{Ostriker2001,heitsch2001,diego2008,soler2013}. Recently, polarization maps have been used to characterize the magnetized turbulence power spectrum within molecular clouds\cite{Hildebrand2009,Houde2011}. These analyses are based on a polarization angle dispersion function and can be used to characterize the relative strength of the ordered and turbulent magnetic field components. The same analysis can be applied to all the BLAST-TNG polarization maps, resulting in a more complete and accurate picture. Also, faint, low column density \textsuperscript{12}CO filaments observed in Taurus\cite{Goldsmith2008,Heyer2008} seem to closely follow magnetic field lines traced by optical polarimetry, which could indicate streaming of molecular gas along field lines. BLAST-TNG will be able to investigate the alignment between the magnetic fields and intermediate and high density filaments seen in \textit{Herschel} maps\cite{Hill2011}. With an increase in knowledge of the role of magnetic fields in molecular clouds, we will come one step closer to understanding how stars form.

\begin{figure}
\centering
\includegraphics[keepaspectratio=true,width=0.6\linewidth]{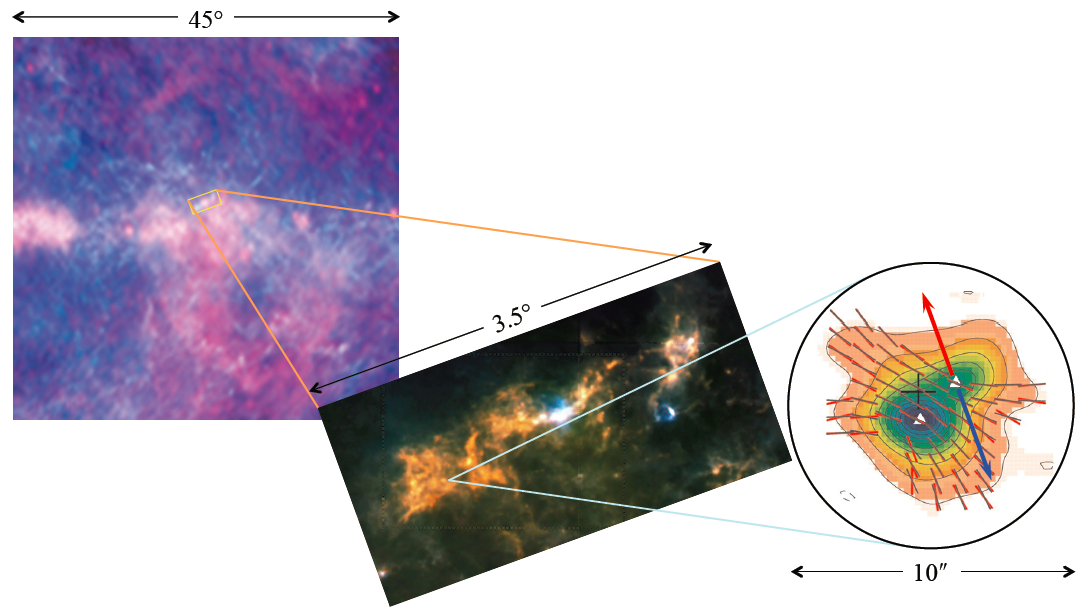}
\caption{BLAST-TNG provides the critical link between \textit{Planck's} all-sky polarimetry at 850~$\mu$m with $5'$ resolution and ALMA's $0.01''$ resolution polarimetry at the same wavelength. The upper left is a Galactic-scale \textit{Planck} image followed by the BLAST observation of Vela\cite{netterfield2009} and the magnetic field map for the IRAS 4A protobinary in Perseus acquired using the Submillimeter Array (a precursor to ALMA)\cite{girart2006}. BLAST-TNG will map polarization using a $22''$ FWHM beam at 250~$\mu$m. This beam nearly matches the ALMA 850~ $\mu$m field-of-view and is more than 200 times smaller (in area) than Planck's 850~$\mu$m beam.}
\label{fig:blastalmalink}
\end{figure}

BLAST-TNG, and its predecessor, BLASTPol, are the first instruments to combine the sensitivity and mapping speed necessary to trace magnetic fields across entire clouds with the resolution to trace fields down into dense substructures, including cores and filaments. BLAST-TNG provides the critical link between the \textit{Planck} all-sky polarization maps with $5'$ resolution and ALMA's ultra-high resolution, narrow ($20''$) field of view\cite{planck1,planck2}. BLAST-TNG will use \textit{Planck}'s all-sky data to refine its target selection. ALMA will then be able to utilize BLASTPol maps to “zero in” on areas of particular interest. Together, these three instruments will be able to probe the inner workings of star formation with previously unreachable resolution, sensitivity and scope.

BLAST was originally designed as an unpolarized instrument to study high redshift dusty infrared galaxies. The conversion to BLASTPol was intended as a proof of concept to see if the technique is viable. While BLASTPol was able to make ground-breaking initial measurements, its utility was limited. In order to make technological progress and fully exploit the targets made available by \textit{Planck}, a next generation instrument powered by large arrays of state-of-the-art detectors must be built. In previous BLASTPol flights, $\sim$5~deg$^{2}$ was mapped in total, while nearby star forming complexes such as Vela or Lupus are 10-100~deg$^{2}$. With BLASTPol we could only map a small fraction of a complex, but with this new instrument, the ability to map the entire complex is readily achievable.

\section{Instrument} \label{sec:instrument}
 This section will focus primarily on components that differ from the previous incarnation of BLASTPol. The design and specifications of this instrument are driven by science goals, availability of existing instrumentation, and the practical limitations of ballooning.
 
\subsection{Optical Design} 

\begin{figure}
\centering
\includegraphics[width=0.8\linewidth]{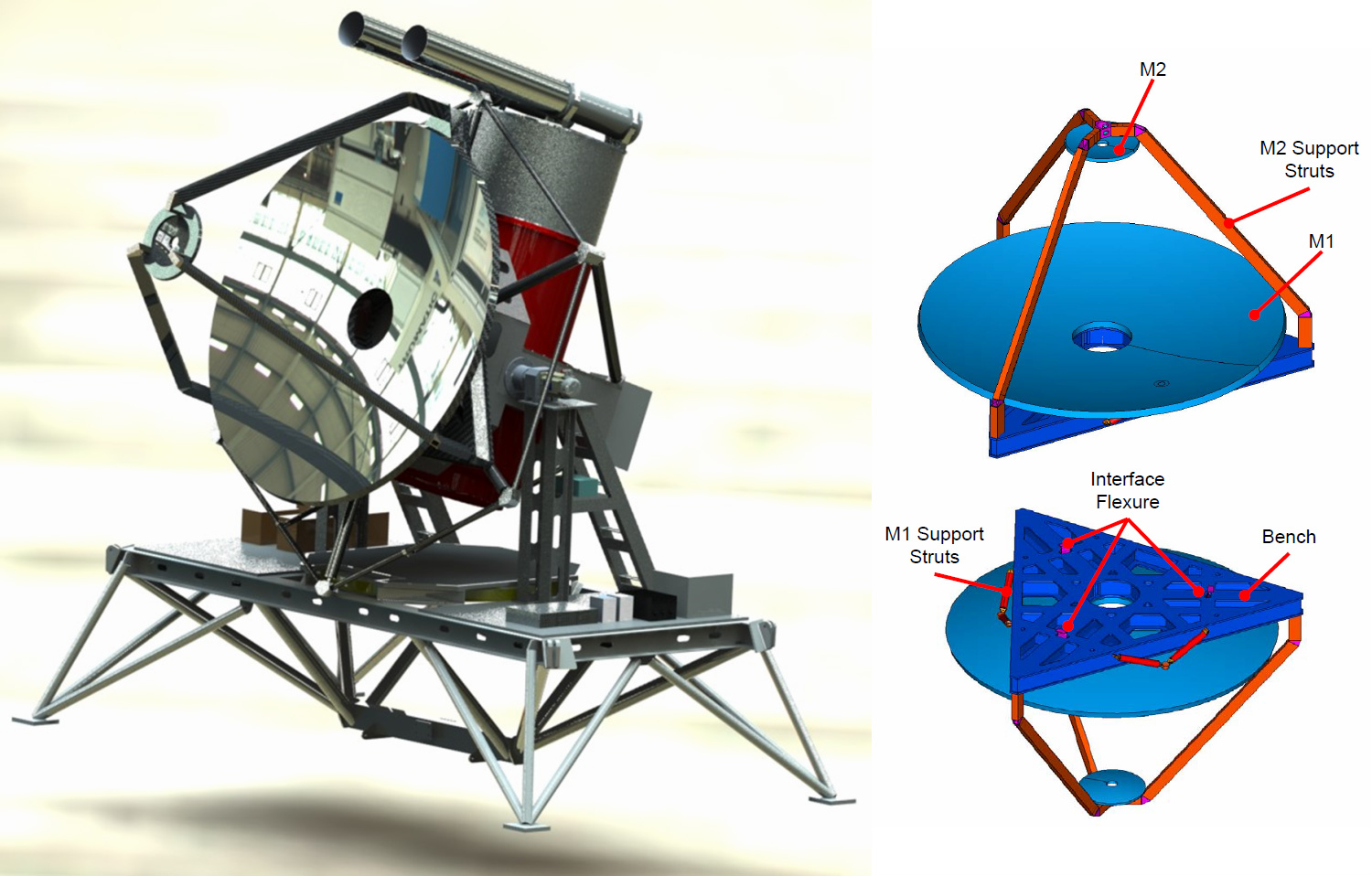}
\caption{Left: The BLAST-TNG gondola design with the inner and outer sun shields removed. Right: The CFRP mirror design concept by Vanguard Space Technologies. The CFRP optics bench supports and constrains both the primary and secondary mirrors at all pointing elevations.}
\label{fig:gondolamirror}
\end{figure}

BLAST-TNG will utilize a 2.5 m carbon fiber reinforced polymer (CFRP) parabolic primary mirror (Figure \ref{fig:gondolamirror}, Right). It will be constructed by Vanguard Space Technologies in San Diego, CA. The carbon fiber primary mirror will be constructed by applying layers of CFRP to a precision-ground positive granite mold. This layer will `sandwich' an aluminum honeycomb support structure. This CFRP / aluminum honeycomb / CFRP sandwich will be attached to a carbon fiber optics bench to maintain its shape at all elevation angles. A 50~cm diameter secondary mirror will be machined out of aluminum and is mounted to the optics bench via three CFRP struts. The secondary mirror will possess an in-flight focusing method via three $\sim$400~g stepper motors. These optics provide diffraction-limited performance over a $22'$ diameter FOV at the Cassegrain focus at $\lambda = 250~\mu$m. The estimated antenna efficiency is $\geq 95\%$. A complete discussion of the previous BLASTPol optics for the 2005 Sweden flight can be found in previous papers\cite{olmi2002}.

\begin{figure}
\centering
\includegraphics[width=0.7\linewidth]{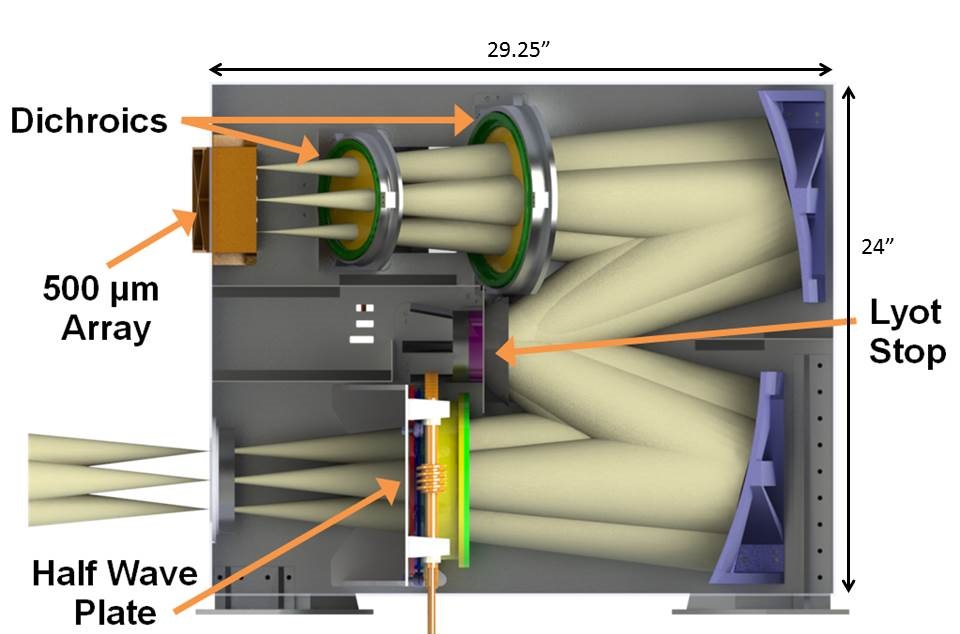}
\caption{A cutaway view of the BLAST-TNG optics box. Light enters from the lower left and is re-imaged onto the three detector arrays (500 $\mu$m array shown at top left). Two dichroic filters split the beam to the 250 and 350~$\mu$m arrays which are mounted on the opposite side of the optics bench. A cold Lyot stop is placed at the image of the primary mirror which allows for greater sidelobe rejection and the introduction of a calibration source. A modulating cold half-wave plate is placed inside the optics box. The motorized mechanism for rotating the waveplate is mounted at 300~K on the inside of the vacuum shell.}
\label{fig:opticsbox}
\end{figure}

The cold reimaging optics (Figure \ref{fig:opticsbox}) are kept at 4~K and consist of an Offner-relay configuration with two concave spherical mirrors and a third cold spherical mirror at the position of an image of the primary mirror. This third spherical mirror acts as a Lyot stop, provides a location for a calibration source, and allows for additional side-lobe rejection. The off-axis angle of the final reimaging mirror allows us to place two dichroic beam splitters in front of the focal plane, making possible simultaneous measurements by all three arrays at different wavelengths.

\subsection{Cryostat Design}   
BLAST-TNG's cryostat design is based on the previous BLAST and BLASTPol cryostat\cite{fissel2010,marsden2008}. Instead of separate LN and LHe reservoirs, the new design calls for a single 250~L LHe tank which forces the boiled-off He gas through two heat exchangers that cool two vapor-cooled shields (VCS), which operate at 66~K and 190~K. We baseline an efficiency of $80\%$ for the heat exchangers, which corresponds to a 28 day hold time, based on testing of a prototype design. We expect to achieve better performance for the fully optimized devices that will be used in the cryostat.

Due to the larger cryogenic volume and proportionally larger optics, the BLAST-TNG cryostat will be ~3.33 times larger in volume than the previous BLAST cryostat. This corresponds to a cryostat diameter of 40 inches and a height of 60 inches. The detectors will be kept at $\sim$280~mK by a closed cycle $^3$He sorption fridge which is backed by a 1~K pumped $^4$He fridge. The self-contained, recycling refrigerator can maintain a base temperature of 280~mK with 30~$\mu$W of cooling power for 4 days. The refrigerator can be recycled in under 2 hours and will be cycled every $\sim$40 hours. The pumped pot maintains 1~K with 20~mW of cooling power with outside atmospheric pressure of 15 Torr or less.

\subsection{Gondola Design} 
The BLAST-TNG gondola provides a pointed platform for the telescope and attaches to the balloon flight train. The gondola consists of two parts: an outer aluminum frame, which can be pointed in azimuth; and an inner aluminum frame which points in elevation. While the BLAST-TNG outer frame remains the same from BLAST and BLASTPol, the gondola inner frame has been completely redesigned to accommodate a $\sim$3.4 times larger cryostat and $40\%$ larger diameter primary mirror (see Figure \ref{fig:gondolamirror}). The pyramids that support the inner frame have been replaced with a step ladder style trusses which are similar to those employed by EBEX\cite{ebex}. 

\begin{figure}
\centering
\includegraphics[width=0.8\linewidth]{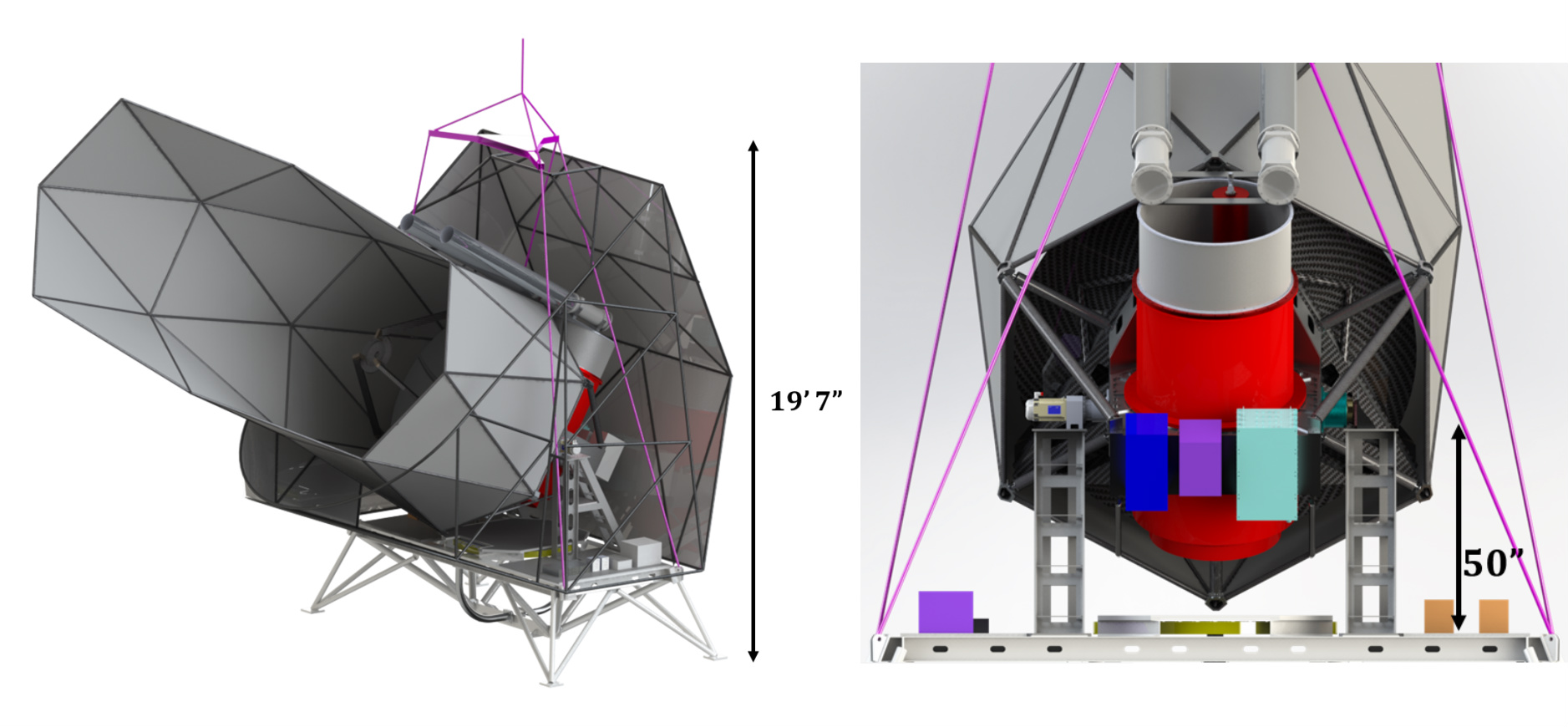}
\caption{The BLAST-TNG sunshield design. The inner baffle allows BLAST-TNG to observe targets between 22-55$^\circ$ elevation and as close as 35$^\circ$ to the sun, while the outer baffle shields the cryostat and electronics from overheating. Left: Right: The rear view of the gondola shows where the readout electronics (teal), power generation (purple) and breakout (blue) hardware enclosures will be mounted on the inner and outer frames.}
\label{fig:nate}
\end{figure}

To avoid large thermal changes in the optics both the inner and outer frames have attached sunshield structures, which is shown in Figure \ref{fig:nate}. In order to accommodate the larger inner frame sunshield and maintain our BLASTPol's maximum pointing elevation of 55~degrees, we also adopt a spreader bar design that is based on one from the EBEX gondola\cite{ebex}.

BLAST-TNG uses the same azimuth control as BLASTPol, which was controlled by a brushless, direct drive servo motor attached to a high moment of inertia reaction wheel, and an active pivot motor which connected the cable suspended gondola to the balloon flight train. The reaction wheel consists of a 1.5~m disk made of 7.6~cm thick aluminum honeycomb, with 48 0.9~kg brass disks mounted around the perimeter. The reaction wheel is mounted below the center of mass of the telescope, directly beneath the active pivot. By spinning the reaction wheel, angular momentum can be transferred to and from the gondola, allowing precise control over the azimuthal velocity of the telescope with minimal latency. The active pivot motor provides additional azimuthal torque by twisting against the flight train, and can also be used over long time scales to transfer angular momentum to the balloon.

BLAST-TNG uses the same suite of sensors (GPS, pinhole sun sensors, magnetometers, gryoscopes, star cameras\cite{rex2006}, and inclinometers) that have been used in previous BLAST and BLASTPol flights. A complete discussion of our hierarchal attitude determination methods and full suite of pointing sensors can be found elsewhere\cite{gandilo2014}.

The elevation of the inner frame is controlled by a Kollmorgon DC brushless servo motor mounted on one side of the inner frame at the attachment point to the outer frame. A free bearing provides the connection point between the inner and outer frames, on the other side. To maintain the dynamic balance of the BLASTPol inner frame, fluid was pumped from a low part of the inner frame to a high part to compensate for the mass lost from cryogen boil-off during flight. However, with the increase in cryogens in BLAST-TNG, the ballast tank system became a higher liability during the flight. Therefore, the servo motor has been fitted with a Harmonic Drive high torque 80:1 gearhead. The Harmonic Drive\footnote{Harmonic Drive LLC. http://harmonicdrive.net/products/gearheads/csg-unit/} has zero backlash and supplies more than enough torque to rotate the inner frame once the cryostat becomes unbalanced after cryogen boil-off. Since the servo motor is now geared, an absolute optical encoder is being used. The RESOLUTE absolute rotary encoder on RESA rings\footnote{Renishaw PLC. http://www.renishaw.com/en/resolute-rotary-angle-absolute-encoder-options--10939} provides resolution to 0.00075~arc-second, which is much more precise than required by BLAST-TNG.

\subsection{Detectors} 
BLAST-TNG will drive the development of polarization-sensitive TiN microwave kinetic inductance detectors (MKIDs) for submillimeter wavelengths. The MKIDs will be divided into three arrays of 627, 324, and 201 feeds at 250, 350, and 500~$\mu$m, respectively (See Table \ref{tab:critnumbers}). Each feedhorn has two orthogonal polarization-sensitive detectors, allowing for simultaneous observations of both polarizations in the same spatial pixel. This design produces just over 2300 detectors in BLAST-TNG, $\sim$8.2 times that of BLASTPol. These new MKID arrays increase the net mapping speed of BLAST-TNG by a factor of 16 over BLASTPol.

\begin{table}
\caption{BLAST-TNG Telescope and receiver parameters}
\centering
\begin{tabular}{cll}
\hline\hline
Telescope: & Temperature & 240~K \\ 
 & Used diameter & 2.4~m (secondary mirror is pupil stopped) \\ 
 & Emissivity & 0.04 \\ 
 &  &  \\ 
Detectors: & Bolometer quantum efficiency & 0.8 \\ 
 & Bolometer feedhorn efficiency & 0.7 \\ 
 & Throughput for each pixel & $A\Omega=\lambda^{2}$ (2.0f$\lambda$ feedhorns fed at f/5)  \\ 
 & Band Centers & \textbf{250 350 500~$\mu$m} \\ 
 & Number of spatial pixels & 627~324~201 \\ 
 & Target Detector $NEP_{freq}$ & 12.3 10.3 6.5~$\times~10^{-17}~W/\sqrt{Hz}$ \\ 
 & Background Power & 17.7 10.6 6.9~pW \\ 
 & Background $NEP_{photon}$ & 16.8 11.0 7.4~$\times~10^{-17}~W/\sqrt{Hz}$ \\  
 &  &  \\ 
Beam: & FWHM & 22 30 42~arc-seconds \\ 
 & Field of view for each array & 22~arc-minute diameter \\ 
 & Overall instrument transmission & $30\%$ \\ 
 & Filter widths $(\lambda/\Delta\lambda)$ & 3 \\ 
 & Observing efficiency & $90\%$ \\ 
 \hline
\end{tabular} 
\label{tab:critnumbers}
\end{table}

The detectors for BLAST-TNG are based on microwave kinetic inductance detectors (MKIDs), a leading detector candidate for future FIR/sub-mm satellite missions. In MKIDs, photons incident on a superconducting film with energy above the gap create a change in quasi-particle density that can be determined by measuring the complex impedance of the film\cite{day2003}. Benefits of MKIDs include built-in frequency domain multiplexing capabilities, the potential for simplified fabrication and the decreased complexity of low temperature cabling and focal plane inter-connects. Thousands of MKIDs, fabricated on a single layer, can be coupled to one RF line and amplified by a low power ($<$5~mW), wide-bandwidth, cryogenic silicon germanium (SiGe) amplifier.

Using MKIDs, a simple design for dual-polarization sensitive detector arrays that is scalable in both pixel count and frequency can be achieved. The detectors are designed to be background-limited in all three BLAST-TNG frequency bands at the predicted in-flight power loads and achieve high efficiency optical coupling. HFSS simulations including orthogonal absorbers, SiOx insulating crossovers, and a quarter wavelength silicon backshort show $>80\%$ co-polar and $<1\%$ cross-polar coupling averaged over the 250 $\mu$m band\cite{hubmayrasc}. More details on our detector design, and MKIDs in general, can be found in other publications\cite{hubmayrltd,vissers2013,doyle2008}.


Optical coupling to the detectors is achieved via an aluminum feedhorn array. The previous BLASTPol feedhorns were the conical \textit{Herschel} SPIRE feedhorn design\cite{spirefeed}. These feedhorns have an asymmetry in X and Y polarizations due to the difference in the E and H fields. This beam asymmetry was a source of cross-polarization in BLASTPol. Therefore, a 3-step modified Potter horn was designed that minimized beam asymmetries and still maintains a 30$\%$ bandwidth\cite{potter,zeng}.

\begin{figure}
\centering
\includegraphics[width=0.7\linewidth]{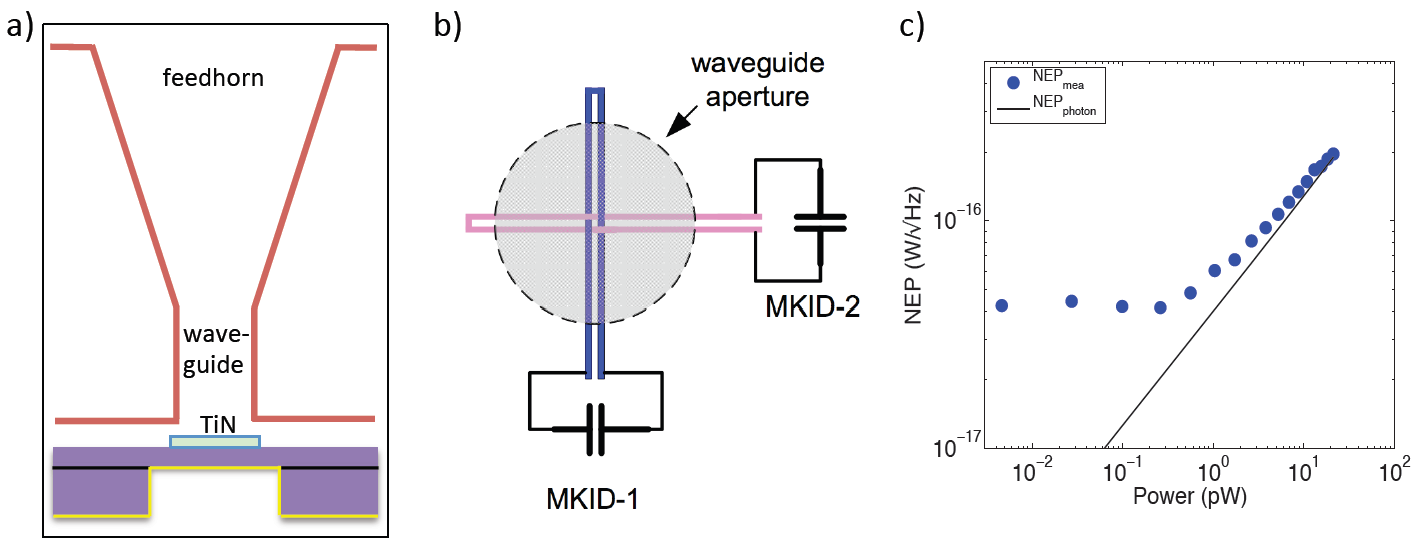}
\caption{Feedhorn-coupled MKID design and performance. \textbf{a)} Cross-section cartoon showing feedhorn-coupling and backshort. \textbf{(b)} Face-on view of two orthogonal MKIDs that achieve dual-polarization sensitivity in the same spatial pixel. \textbf{(c)} BLAST-TNG prototype detector measured NEP versus thermal load. The device achieves background limited performance for loading $>$ 1~pW.}
\label{fig:detectors}
\end{figure}

In order to verify that the detectors will achieve the desired noise performance, a prototype subarray of detectors was fabricated and tested with a variable temperature blackbody load. These tests, shown in Figure \ref{fig:detectors}(c), show photon-noise limited sensitivity over a broad range of loading conditions that includes the predicted in-flight load for each band. Further work is in progress to test the detector's noise level in our scan speed governed science band of 0.01-0.1~Hz, and will be discussed in an upcoming publication.

\begin{figure}
\centering
\includegraphics[width=0.7\linewidth]{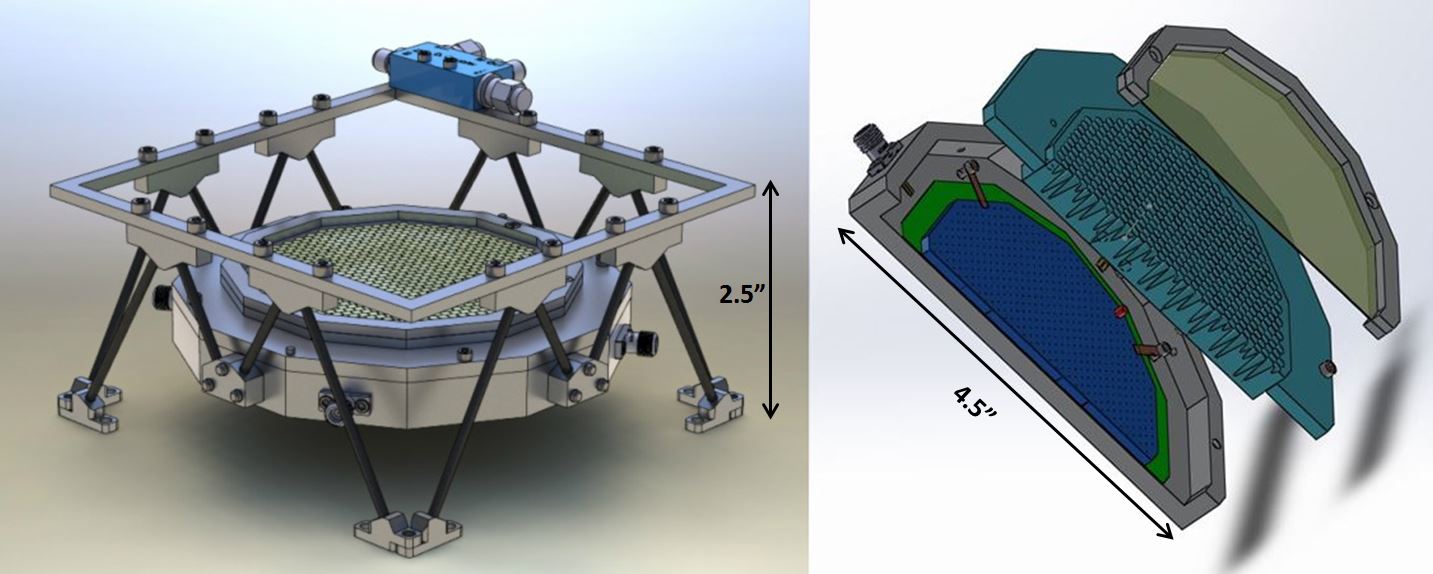}
\caption{Focal Plane Array (FPA) design and assembly. Left: A rendering of the FPA in its two-stage CFRP thermal standoff. The top ring sits at 1~K and provides a mount point for the directional coupler while the FPA sits at 280~mK. Right: A cross-sectional exploded view of the FPA. The detector wafer (green) is mounted directly below a square waveguide interface wafer (blue). Both of these elements are fixed in X/Y via a pin (orange) and slot method, and fixed in Z via several beryllium copper spring tabs. The feedhorns (teal) are mounted on top, as well as a bandpass filter.}
\label{fig:fpa_combo}
\end{figure}

The detectors are housed in a focal plane array housing (FPA) which is an 8-sided polygon and is shown in Figure \ref{fig:fpa_combo}. The backplate possesses a rim for mounting the thermal standoff, SMA connectors, and heat strap to the 300~mK absorption fridge. The detector and square waveguide interface wafers are mounted inside the backplate via a pin-and-slot method to constrain the detectors in X/Y. To fix the Z dimension, both wafers are held down to the backplate via beryllium copper spring tabs. The detector wafer is oversized compared to the square interface wafer to allow for wirebonds to both the microstrip multiplexing lines. Additionally, gold wirebonds are made to increase the thermal conductivity and keep the detectors cooled to 280~mK. Finally, the feedhorn plate attaches to the outer rim of the backplate and is constrained via two close-fit pins. A bandpass filter is attached to the front of the feedhorn plate.

The thermal standoff consists of two interlocking sets of thin-walled CFRP rods. The first set of rods are 2 inches long and isolate the 4 K optics box to a ring at 1~K. One set of SMA cables is heat sunk to the 1~K stage via a directional coupler. Next, a set of one inch long rods isolate the 1~K stage from the FPA which is cooled to 280~mK. Thermal conductivity calculations show that the load to the 280~mK and 1~K stages will be $\sim$14~$\mu$W and $\sim$ 120~$\mu$W, respectively, for all three arrays. Finite element analysis on the carbon fiber support structure shows $<5~\mu$m of deflection during normal operation. This amount of thermal loading and deflection is well within our tolerances. 

\subsection{Polarization Modulation}
The primary modulation scheme for BLAST-TNG will be based on the successful implementation of a broad-band half-wave plate (HWP) used in BLASTPol shown in Figure \ref{fig:opticsbox}. The low-emissivity HWP consists of five 500~$\mu$m sapphire layers glued together with 6~$\mu$m layers of polyethylene. The HWP has a metal mesh anti-reflective coating\cite{zhang2009} with a measured in band transmission of 86-95$\%$. Located inside the cryogenic optics box, the HWP is far from the Cassegrain focus ($\sim$30~cm) while allowing it to dissipate heat to the 4~K stage. Raster maps with areas ranging from 0.25 to tens of square degrees are made by scanning the gondola in azimuth and elevation. The HWP orientation is changed at the end of each raster scan ($\sim$10 min). A typical 5 hour integration therefore consists of $\sim$30 HWP moves. The integration is spread over the flight to maximize the sky rotation while taking full advantage of the polarization modulation.

BLAST-TNG has two orthogonally polarized detectors per pixel with the vectors rotated by 45$\deg$ from pixel to pixel along the scan direction. As the telescope scans at a typical speed of $0.1^\circ$/s, this configuration provides near simultaneous measurements of both Q and U while eliminating most of the common-mode 1/f noise. It also provides a ``fail safe" backup should the HWP rotation mechanism somehow fail. The BLAST-TNG cold optics are larger, similar versions of what was used for BLASTPol (Figure \ref{fig:opticsbox}), which has a measured instrumental polarization of $< 0.5\%$. The symmetry of our on-axis telescope provides low inherent instrumental polarization. Operation in the submillimeter carries the risk of diffraction and scattering off of the secondary mirror and its supports. However, there was no evidence of this with BLASTPol. The signals from the dust polarization are much larger in the submillimeter than they are in the CMB bands. Therefore, BLAST-TNG does not require the extreme control of systematic effects that are needed to measure CMB polarization directly.

\acknowledgments     
 
The BLAST collaboration acknowledges the support of NASA through grant numbers NNX13AE50G S04 and NNX09AB98G, the Leverhulme Trust through the Research Project Grant F/00 407/BN, Canada's Natural Sciences and Engineering Research Council (NSERC), and the National Science Foundation Office of Polar Programs. Frederick Poidevin thanks the Spanish Ministry of Economy and Competitiveness (MINECO) under the Consolider-Ingenio project CSD2010-00064 (EPI: Exploring the Physics of Inflation). Bradley Dober is supported through a NASA Earth and Space Science Fellowship. Peter Ashton is supported through \textit{Reach for the Stars}, a GK-12 program supported by the National Science Foundation under grant DGE-0948017. However, any opinions, findings, conclusions, and/or recommendations are those of the investigators and do not necessarily reflect the views of the Foundation. We would also like to thank the Columbia Scientific Balloon Facility (CSBF) staff for their continued outstanding work.


\bibliography{report}   

\begin{thebibliography}{10}

\bibitem{matthews2013}
Matthews, T., Ade, P., Angil{\`e}, F., Benton, S., Chapman, N., Devlin, M.,
  Dober, B., Fissel, L., Fukui, Y., Gandilo, N., et~al., ``2010 blastpol
  observations of the magnetic field of the filamentary galactic cloud 'lupus
  i','' in [{\em American Astronomical Society Meeting
  Abstracts}{\nolinebreak\hspace{0.1em}]},   {\bf 222} (2013).

\bibitem{poidevin2014}
Poidevin, F., Ade, P., Angile, F., Benton, S., Chapin, E., Devlin, M., Fissel,
  L., Fukui, Y., Gandilo, N., Gundersen, J., et~al., ``Comparison of prestellar
  core elongations and large-scale molecular cloud structures in the lupus i
  region,'' {\em arXiv preprint arXiv:1405.0331}  (2014).

\bibitem{galitzki2014}
{Galitzki}, N., {Ade}, P. A.~R., {Angil\`e}, F.~E., {Benton}, S.~J., {Devlin},
  M.~J., {Dober}, B., {Fissel}, L.~M., {Fukui}, Y., {Gandilo}, N.~N., {Klein},
  J., {Korotkov}, A.~L., {Matthews}, T.~G., {Moncelsi}, L., {Netterfield},
  C.~B., {Novak}, G., {Nutter}, D., {Pascale}, E., {Poidevin}, F., {Savini},
  G., {Scott}, D., {Shariff}, J.~A., {Soler}, J.~D., {Tucker}, C.~E., {Tucker},
  G.~S., and {Ward-Thompson}, D., ``{The Balloon-borne Large Aperture
  Submillimeter Telescope for Polarimetry-BLASTPol: Performance and results
  from the 2012 Antarctic flight},'' in [{\em Ground-based and Airborne
  Telescopes V}{\nolinebreak\hspace{0.1em}]},  {\em Presented at the Society of
  Photo-Optical Instrumentation Engineers (SPIE) Conference} {\bf 9145} (June
  2014).

\bibitem{Hennebelle2011}
{Hennebelle}, P., {Commer{\c c}on}, B., {Joos}, M., {Klessen}, R.~S.,
  {Krumholz}, M., {Tan}, J.~C., and {Teyssier}, R., ``{Collapse, outflows and
  fragmentation of massive, turbulent and magnetized prestellar barotropic
  cores},'' {\em \aap}~{\bf 528},  A72 (Apr. 2011).

\bibitem{LiNak2004}
Li, Z.-Y. and Nakamura, F., ``Magnetically regulated star formation in
  turbulent clouds,'' {\em The Astrophysical Journal Letters}~{\bf 609}(2),
  L83 (2004).

\bibitem{Li2010}
Li, Z.-Y., Wang, P., Abel, T., and Nakamura, F., ``Lowering the characteristic
  mass of cluster stars by magnetic fields and outflow feedback,'' {\em The
  Astrophysical Journal Letters}~{\bf 720}(1),  L26 (2010).

\bibitem{Dotson2010}
Dotson, J.~L., Vaillancourt, J.~E., Kirby, L., Dowell, C.~D., Hildebrand,
  R.~H., and Davidson, J.~A., ``350 $\mu$m polarimetry from the caltech
  submillimeter observatory,'' {\em The Astrophysical Journal Supplement
  Series}~{\bf 186}(2),  406 (2010).

\bibitem{Matthews2009}
Matthews, B.~C., McPhee, C.~A., Fissel, L.~M., and Curran, R.~L., ``The legacy
  of scupol: 850 $\mu$m imaging polarimetry from 1997 to 2005,'' {\em The
  Astrophysical Journal Supplement Series}~{\bf 182}(1),  143 (2009).

\bibitem{Ostriker2001}
Ostriker, E.~C., Stone, J.~M., and Gammie, C.~F., ``Density, velocity, and
  magnetic field structure in turbulent molecular cloud models,'' {\em The
  Astrophysical Journal}~{\bf 546}(2),  980 (2001).

\bibitem{heitsch2001}
{Heitsch}, F., {Zweibel}, E.~G., {Mac Low}, M.-M., {Li}, P., and {Norman},
  M.~L., ``{Magnetic Field Diagnostics Based on Far-Infrared Polarimetry: Tests
  Using Numerical Simulations},'' {\em \apj}~{\bf 561},  800--814 (Nov. 2001).

\bibitem{diego2008}
{Falceta-Gon{\c c}alves}, D., {Lazarian}, A., and {Kowal}, G., ``{Studies of
  Regular and Random Magnetic Fields in the ISM: Statistics of Polarization
  Vectors and the Chandrasekhar-Fermi Technique},'' {\em \apj}~{\bf 679},
  537--551 (May 2008).

\bibitem{soler2013}
{Soler}, J.~D., {Hennebelle}, P., {Martin}, P.~G., {Miville-Desch{\^e}nes},
  M.-A., {Netterfield}, C.~B., and {Fissel}, L.~M., ``{An Imprint of Molecular
  Cloud Magnetization in the Morphology of the Dust Polarized Emission},'' {\em
  \apj}~{\bf 774},  128 (Sept. 2013).

\bibitem{Hildebrand2009}
Hildebrand, R.~H., Kirby, L., Dotson, J.~L., Houde, M., and Vaillancourt,
  J.~E., ``Dispersion of magnetic fields in molecular clouds. i,'' {\em The
  Astrophysical Journal}~{\bf 696}(1),  567 (2009).

\bibitem{Houde2011}
Houde, M., Rao, R., Vaillancourt, J.~E., and Hildebrand, R.~H., ``Dispersion of
  magnetic fields in molecular clouds. iii.,'' {\em The Astrophysical
  Journal}~{\bf 733}(2),  109 (2011).

\bibitem{Goldsmith2008}
Goldsmith, P.~F., Heyer, M., Narayanan, G., Snell, R., Li, D., and Brunt, C.,
  ``Large-scale structure of the molecular gas in taurus revealed by high
  linear dynamic range spectral line mapping,'' {\em The Astrophysical
  Journal}~{\bf 680}(1),  428 (2008).

\bibitem{Heyer2008}
Heyer, M., Gong, H., Ostriker, E., and Brunt, C., ``Magnetically aligned
  velocity anisotropy in the taurus molecular cloud,'' {\em The Astrophysical
  Journal}~{\bf 680}(1),  420 (2008).

\bibitem{Hill2011}
{Hill}, T., {Motte}, F., {Didelon}, P., {Bontemps}, S., {Minier}, V.,
  {Hennemann}, M., {Schneider}, N., {Andr{\'e}}, P., {Men'shchikov}, A.,
  {Anderson}, L.~D., {Arzoumanian}, D., {Bernard}, J.-P., {di Francesco}, J.,
  {Elia}, D., {Giannini}, T., {Griffin}, M.~J., {K{\"o}nyves}, V., {Kirk}, J.,
  {Marston}, A.~P., {Martin}, P.~G., {Molinari}, S., {Nguyen Luong}, Q.,
  {Peretto}, N., {Pezzuto}, S., {Roussel}, H., {Sauvage}, M., {Sousbie}, T.,
  {Testi}, L., {Ward-Thompson}, D., {White}, G.~J., {Wilson}, C.~D., and
  {Zavagno}, A., ``{Filaments and ridges in Vela C revealed by Herschel: from
  low-mass to high-mass star-forming sites},'' {\em \aap}~{\bf 533},  A94
  (Sept. 2011).

\bibitem{netterfield2009}
Netterfield, C.~B., Ade, P.~A., Bock, J.~J., Chapin, E.~L., Devlin, M.~J.,
  Griffin, M., Gundersen, J.~O., Halpern, M., Hargrave, P.~C., Hughes, D.~H.,
  et~al., ``Blast: The mass function, lifetimes, and properties of intermediate
  mass cores from a 50 deg2 submillimeter galactic survey in vela (ℓ
  265°),'' {\em The Astrophysical Journal}~{\bf 707}(2),  1824 (2009).

\bibitem{girart2006}
Girart, J.~M., Rao, R., and Marrone, D.~P., ``Magnetic fields in the formation
  of sun-like stars,'' {\em Science}~{\bf 313}(5788),  812--814 (2006).

\bibitem{planck1}
{Planck Collaboration}, {Ade}, P.~A.~R., {Aghanim}, N., {Alves}, M.~I.~R.,
  {Aniano}, G., {Armitage-Caplan}, C., {Arnaud}, M., {Arzoumanian}, D.,
  {Ashdown}, M., {Atrio-Barandela}, F., {Aumont}, J., {Baccigalupi}, C.,
  {Banday}, A.~J., {Barreiro}, R.~B., {Battaner}, E., {Benabed}, K.,
  {Benoit-L{\'e}vy}, A., {Bernard}, J.-P., {Bersanelli}, M., {Bielewicz}, P.,
  {Bond}, J.~R., {Borrill}, J., {Bouchet}, F.~R., {Boulanger}, F., {Bracco},
  A., {Burigana}, C., {Cardoso}, J.-F., {Catalano}, A., {Chamballu}, A.,
  {Chiang}, H.~C., {Christensen}, P.~R., {Colombi}, S., {Colombo}, L.~P.~L.,
  {Combet}, C., {Couchot}, F., {Coulais}, A., {Crill}, B.~P., {Curto}, A.,
  {Cuttaia}, F., {Danese}, L., {Davies}, R.~D., {Davis}, R.~J., {de Bernardis},
  P., {de Rosa}, A., {de Zotti}, G., {Delabrouille}, J., {Dickinson}, C.,
  {Diego}, J.~M., {Donzelli}, S., {Dor{\'e}}, O., {Douspis}, M., {Dupac}, X.,
  {En{\ss}lin}, T.~A., {Eriksen}, H.~K., {Falgarone}, E., {Fanciullo}, L.,
  {Ferri{\`e}re}, K., {Finelli}, F., {Forni}, O., {Frailis}, M., {Fraisse},
  A.~A., {Franceschi}, E., {Galeotta}, S., {Ganga}, K., {Ghosh}, T., {Giard},
  M., {Giraud-H{\'e}raud}, Y., {Gonz{\'a}lez-Nuevo}, J., {G{\'o}rski}, K.~M.,
  {Gregorio}, A., {Gruppuso}, A., {Guillet}, V., {Hansen}, F.~K., {Harrison},
  D.~L., {Helou}, G., {Hern{\'a}ndez-Monteagudo}, C., {Hildebrandt}, S.~R.,
  {Hivon}, E., {Hobson}, M., {Holmes}, W.~A., {Hornstrup}, A., {Huffenberger},
  K.~M., {Jaffe}, A.~H., {Jaffe}, T.~R., {Jones}, W.~C., {Juvela}, M.,
  {Keih{\"a}nen}, E., {Keskitalo}, R., {Kisner}, T.~S., {Kneissl}, R.,
  {Knoche}, J., {Kunz}, M., {Kurki-Suonio}, H., {Lagache}, G., {Lamarre},
  J.-M., {Lasenby}, A., {Lawrence}, C.~R., {Leonardi}, R., {Levrier}, F.,
  {Liguori}, M., {Lilje}, P.~B., {Linden-V{\o}rnle}, M., {L{\'o}pez-Caniego},
  M., {Lubin}, P.~M., {Mac{\'{\i}}as-P{\'e}rez}, J.~F., {Maino}, D.,
  {Mandolesi}, N., {Maris}, M., {Marshall}, D.~J., {Martin}, P.~G.,
  {Mart{\'{\i}}nez-Gonz{\'a}lez}, E., {Masi}, S., {Matarrese}, S., {Mazzotta},
  P., {Melchiorri}, A., {Mendes}, L., {Mennella}, A., {Migliaccio}, M.,
  {Miville-Desch{\^e}nes}, M.-A., {Moneti}, A., {Montier}, L., {Morgante}, G.,
  {Mortlock}, D., {Munshi}, D., {Murphy}, J.~A., {Naselsky}, P., {Nati}, F.,
  {Natoli}, P., {Netterfield}, C.~B., {Noviello}, F., {Novikov}, D., {Novikov},
  I., {Oxborrow}, C.~A., {Pagano}, L., {Pajot}, F., {Paoletti}, D., {Pasian},
  F., {Pelkonen}, V.-M., {Perdereau}, O., {Perotto}, L., {Perrotta}, F.,
  {Piacentini}, F., {Piat}, M., {Pietrobon}, D., {Plaszczynski}, S.,
  {Pointecouteau}, E., {Polenta}, G., {Popa}, L., {Pratt}, G.~W., {Prunet}, S.,
  {Puget}, J.-L., {Rachen}, J.~P., {Reinecke}, M., {Remazeilles}, M.,
  {Renault}, C., {Ricciardi}, S., {Riller}, T., {Ristorcelli}, I., {Rocha}, G.,
  {Rosset}, C., {Roudier}, G., {Rusholme}, B., {Sandri}, M., {Scott}, D.,
  {Soler}, J.~D., {Spencer}, L.~D., {Stolyarov}, V., {Stompor}, R., {Sudiwala},
  R., {Sutton}, D., {Suur-Uski}, A.-S., {Sygnet}, J.-F., {Tauber}, J.~A.,
  {Terenzi}, L., {Toffolatti}, L., {Tomasi}, M., {Tristram}, M., {Tucci}, M.,
  {Umana}, G., {Valenziano}, L., {Valiviita}, J., {Van Tent}, B., {Vielva}, P.,
  {Villa}, F., {Wade}, L.~A., {Wandelt}, B.~D., and {Zonca}, A., ``{Planck
  intermediate results. XIX. An overview of the polarized thermal emission from
  Galactic dust},'' {\em ArXiv e-prints}  (May 2014).

\bibitem{planck2}
{Planck Collaboration}, {Ade}, P.~A.~R., {Aghanim}, N., {Alves}, M.~I.~R.,
  {Aniano}, G., {Armitage-Caplan}, C., {Arnaud}, M., {Arzoumanian}, D.,
  {Ashdown}, M., {Atrio-Barandela}, F., {Aumont}, J., {Baccigalupi}, C.,
  {Banday}, A.~J., {Barreiro}, R.~B., {Battaner}, E., {Benabed}, K.,
  {Benoit-L{\'e}vy}, A., {Bernard}, J.-P., {Bersanelli}, M., {Bielewicz}, P.,
  {Bond}, J.~R., {Borrill}, J., {Bouchet}, F.~R., {Boulanger}, F., {Bracco},
  A., {Burigana}, C., {Cardoso}, J.-F., {Catalano}, A., {Chamballu}, A.,
  {Chiang}, H.~C., {Christensen}, P.~R., {Colombi}, S., {Colombo}, L.~P.~L.,
  {Combet}, C., {Couchot}, F., {Coulais}, A., {Crill}, B.~P., {Curto}, A.,
  {Cuttaia}, F., {Danese}, L., {Davies}, R.~D., {Davis}, R.~J., {de Bernardis},
  P., {de Rosa}, A., {de Zotti}, G., {Delabrouille}, J., {Dickinson}, C.,
  {Diego}, J.~M., {Donzelli}, S., {Dor{\'e}}, O., {Douspis}, M., {Dupac}, X.,
  {En{\ss}lin}, T.~A., {Eriksen}, H.~K., {Falgarone}, E., {Fanciullo}, L.,
  {Ferri{\`e}re}, K., {Finelli}, F., {Forni}, O., {Frailis}, M., {Fraisse},
  A.~A., {Franceschi}, E., {Galeotta}, S., {Ganga}, K., {Ghosh}, T., {Giard},
  M., {Giraud-H{\'e}raud}, Y., {Gonz{\'a}lez-Nuevo}, J., {G{\'o}rski}, K.~M.,
  {Gregorio}, A., {Gruppuso}, A., {Guillet}, V., {Hansen}, F.~K., {Harrison},
  D.~L., {Helou}, G., {Hern{\'a}ndez-Monteagudo}, C., {Hildebrandt}, S.~R.,
  {Hivon}, E., {Hobson}, M., {Holmes}, W.~A., {Hornstrup}, A., {Huffenberger},
  K.~M., {Jaffe}, A.~H., {Jaffe}, T.~R., {Jones}, W.~C., {Juvela}, M.,
  {Keih{\"a}nen}, E., {Keskitalo}, R., {Kisner}, T.~S., {Kneissl}, R.,
  {Knoche}, J., {Kunz}, M., {Kurki-Suonio}, H., {Lagache}, G., {Lamarre},
  J.-M., {Lasenby}, A., {Lawrence}, C.~R., {Leonardi}, R., {Levrier}, F.,
  {Liguori}, M., {Lilje}, P.~B., {Linden-V{\o}rnle}, M., {L{\'o}pez-Caniego},
  M., {Lubin}, P.~M., {Mac{\'{\i}}as-P{\'e}rez}, J.~F., {Maino}, D.,
  {Mandolesi}, N., {Maris}, M., {Marshall}, D.~J., {Martin}, P.~G.,
  {Mart{\'{\i}}nez-Gonz{\'a}lez}, E., {Masi}, S., {Matarrese}, S., {Mazzotta},
  P., {Melchiorri}, A., {Mendes}, L., {Mennella}, A., {Migliaccio}, M.,
  {Miville-Desch{\^e}nes}, M.-A., {Moneti}, A., {Montier}, L., {Morgante}, G.,
  {Mortlock}, D., {Munshi}, D., {Murphy}, J.~A., {Naselsky}, P., {Nati}, F.,
  {Natoli}, P., {Netterfield}, C.~B., {Noviello}, F., {Novikov}, D., {Novikov},
  I., {Oxborrow}, C.~A., {Pagano}, L., {Pajot}, F., {Paoletti}, D., {Pasian},
  F., {Pelkonen}, V.-M., {Perdereau}, O., {Perotto}, L., {Perrotta}, F.,
  {Piacentini}, F., {Piat}, M., {Pietrobon}, D., {Plaszczynski}, S.,
  {Pointecouteau}, E., {Polenta}, G., {Popa}, L., {Pratt}, G.~W., {Prunet}, S.,
  {Puget}, J.-L., {Rachen}, J.~P., {Reinecke}, M., {Remazeilles}, M.,
  {Renault}, C., {Ricciardi}, S., {Riller}, T., {Ristorcelli}, I., {Rocha}, G.,
  {Rosset}, C., {Roudier}, G., {Rusholme}, B., {Sandri}, M., {Scott}, D.,
  {Soler}, J.~D., {Spencer}, L.~D., {Stolyarov}, V., {Stompor}, R., {Sudiwala},
  R., {Sutton}, D., {Suur-Uski}, A.-S., {Sygnet}, J.-F., {Tauber}, J.~A.,
  {Terenzi}, L., {Toffolatti}, L., {Tomasi}, M., {Tristram}, M., {Tucci}, M.,
  {Umana}, G., {Valenziano}, L., {Valiviita}, J., {Van Tent}, B., {Vielva}, P.,
  {Villa}, F., {Wade}, L.~A., {Wandelt}, B.~D., and {Zonca}, A., ``{Planck
  intermediate results. XX. Comparison of polarized thermal emission from
  Galactic dust with simulations of MHD turbulence},'' {\em ArXiv e-prints}
  (May 2014).

\bibitem{olmi2002}
Olmi, L., ``Optical designs for submillimeter-wave spherical-primary(sub)
  orbital telescopes and novel optimization techniques,'' in [{\em Proceedings
  of SPIE}{\nolinebreak\hspace{0.1em}]},   {\bf 4849},  245--256 (2002).

\bibitem{fissel2010}
{Fissel}, L.~M., {Ade}, P.~A.~R., {Angil{\`e}}, F.~E., {Benton}, S.~J.,
  {Chapin}, E.~L., {Devlin}, M.~J., {Gandilo}, N.~N., {Gundersen}, J.~O.,
  {Hargrave}, P.~C., {Hughes}, D.~H., {Klein}, J., {Korotkov}, A.~L.,
  {Marsden}, G., {Matthews}, T.~G., {Moncelsi}, L., {Mroczkowski}, T.~K.,
  {Netterfield}, C.~B., {Novak}, G., {Olmi}, L., {Pascale}, E., {Savini}, G.,
  {Scott}, D., {Shariff}, J.~A., {Soler}, J.~D., {Thomas}, N.~E., {Truch},
  M.~D.~P., {Tucker}, C.~E., {Tucker}, G.~S., {Ward-Thompson}, D., and {Wiebe},
  D.~V., ``{The balloon-borne large-aperture submillimeter telescope for
  polarimetry: BLAST-Pol},'' in [{\em Society of Photo-Optical Instrumentation
  Engineers (SPIE) Conference Series}{\nolinebreak\hspace{0.1em}]},  {\em
  Society of Photo-Optical Instrumentation Engineers (SPIE) Conference Series}
  {\bf 7741} (July 2010).

\bibitem{marsden2008}
{Marsden}, G., {Ade}, P.~A.~R., {Benton}, S., {Bock}, J.~J., {Chapin}, E.~L.,
  {Chung}, J., {Devlin}, M.~J., {Dicker}, S., {Fissel}, L., {Griffin}, M.,
  {Gundersen}, J.~O., {Halpern}, M., {Hargrave}, P.~C., {Hughes}, D.~H.,
  {Klein}, J., {Korotkov}, A., {MacTavish}, C.~J., {Martin}, P.~G., {Martin},
  T.~G., {Matthews}, T.~G., {Mauskopf}, P., {Moncelsi}, L., {Netterfield},
  C.~B., {Novak}, G., {Pascale}, E., {Olmi}, L., {Patanchon}, G., {Rex}, M.,
  {Savini}, G., {Scott}, D., {Semisch}, C., {Thomas}, N., {Truch}, M.~D.~P.,
  {Tucker}, C., {Tucker}, G.~S., {Viero}, M.~P., {Ward-Thompson}, D., and
  {Wiebe}, D.~V., ``{The Balloon-borne Large-Aperture Submillimeter Telescope
  for polarization: BLAST-pol},'' in [{\em Society of Photo-Optical
  Instrumentation Engineers (SPIE) Conference
  Series}{\nolinebreak\hspace{0.1em}]},  {\em Society of Photo-Optical
  Instrumentation Engineers (SPIE) Conference Series} {\bf 7020} (Aug. 2008).

\bibitem{ebex}
Reichborn-Kjennerud, B., Aboobaker, A.~M., Ade, P., Aubin, F., Baccigalupi, C.,
  Bao, C., Borrill, J., Cantalupo, C., Chapman, D., Didier, J., et~al., ``Ebex:
  a balloon-borne cmb polarization experiment,'' in [{\em SPIE Astronomical
  Telescopes+ Instrumentation}{\nolinebreak\hspace{0.1em}]},   77411C--77411C,
  International Society for Optics and Photonics (2010).

\bibitem{rex2006}
Rex, M., Chapin, E., Devlin, M.~J., Gundersen, J., Klein, J., Pascale, E., and
  Wiebe, D., ``Blast autonomous daytime star cameras,'' in [{\em Astronomical
  Telescopes and Instrumentation}{\nolinebreak\hspace{0.1em}]},
  62693H--62693H, International Society for Optics and Photonics (2006).

\bibitem{gandilo2014}
{Gandilo}, N.~N., {Ade}, P.~A.~R., {Amiri}, M., {Angil\`e}, F.~E., {Benton},
  S.~J., {Bock}, J.~J., {Bond}, J.~R., {Bonetti}, J.~A., {Bryan}, S.~A.,
  {Chiang}, H.~C., {Contaldi}, C.~R., {Crill}, B.~P., {Devlin}, M.~J., {Dober},
  B., {Dor\'e}, O.~P., {Farhang}, M., {Filippini}, J.~P., {Fissel}, L.~M.,
  {Fraisse}, A.~A., {Fukui}, Y., {Galitzki}, N., {Gambrel}, A.~E.,
  {Gudmundsson}, J.~E., {Halpern}, M., {Hasselfield}, M., {Hilton}, G.~C.,
  {Holmes}, W.~A., {Hristov}, V.~V., {Irwin}, K.~D., {Jones}, W.~C., {Kermish},
  Z.~D., {Klein}, J., {Korotkov}, A.~L., {MacTavish}, C.~J., {Mason}, P.~V.,
  {Matthews}, T.~G., {Megerian}, K.~G., {Moncelsi}, L., {Montroy}, T.~E.,
  {Morford}, T.~A., {Mroczkowski}, T.~K., {Nagy}, J.~M., {Netterfield}, C.~B.,
  {Novak}, G., {Nutter}, D., {Pascale}, E., {Poidevin}, F., {Rahlin}, A.~S.,
  {Reintsema}, C.~D., {Ruhl}, J.~E., {Runyan}, M.~C., {Savini}, G., {Scott},
  D., {Shariff}, J.~A., {Soler}, J.~D., {Thomas}, N.~E., {Trangsrud}, A.,
  {Truch}, M.~D., {Tucker}, C.~E., {Tucker}, G.~S., {Tucker}, R.~S., {Turner},
  A.~D., {Ward-Thompson}, D., {Weber}, A.~C., {Wiebe}, D.~V., and {Young},
  E.~Y., ``{Attitude determination for balloon-borne experiments},'' in [{\em
  Ground-based and Airborne Telescopes V}{\nolinebreak\hspace{0.1em}]},  {\em
  Presented at the Society of Photo-Optical Instrumentation Engineers (SPIE)
  Conference} {\bf 9145-101} (June 2014).

\bibitem{day2003}
Day, P.~K., LeDuc, H.~G., Mazin, B.~A., Vayonakis, A., and Zmuidzinas, J., ``A
  broadband superconducting detector suitable for use in large arrays,'' {\em
  Nature}~{\bf 425}(6960),  817--821 (2003).

\bibitem{hubmayrasc}
Hubmayr, J., Beall, J., Becker, D., Cho, H., Dober, B., Devlin, M., Fox, A.,
  Gao, J., Hilton, G., Irwin, K., et~al., ``Dual-polarization sensitive mkids
  for far infrared astrophysics,'' {\em Applied Superconductivity, IEEE
  Transactions on}~{\bf 23}(3),  2400304--2400304 (2013).

\bibitem{hubmayrltd}
Hubmayr, J., Beall, J., Becker, D., Brevik, J., Cho, H., Che, G., Devlin, M.,
  Dober, B., Gao, J., Galitzki, N., et~al., ``Dual-polarization-sensitive
  kinetic inductance detectors for balloon-borne sub-millimeter polarimetry,''
  {\em Journal of Low Temperature Physics} ,  1--7 (2014).

\bibitem{vissers2013}
Vissers, M.~R., Gao, J., Sandberg, M., Duff, S.~M., Wisbey, D.~S., Irwin,
  K.~D., and Pappas, D.~P., ``Proximity-coupled ti/tin multilayers for use in
  kinetic inductance detectors,'' {\em Applied Physics Letters}~{\bf 102}(23),
  232603 (2013).

\bibitem{doyle2008}
Doyle, S., Mauskopf, P., Naylon, J., Porch, A., and Duncombe, C., ``Lumped
  element kinetic inductance detectors,'' {\em Journal of Low Temperature
  Physics}~{\bf 151}(1-2),  530--536 (2008).

\bibitem{spirefeed}
Chattopadhyay, G., Glenn, J., Bock, J.~J., Rownd, B.~K., Caldwell, M., and
  Griffin, M.~J., ``Feed horn coupled bolometer arrays for spire-design,
  simulations, and measurements,'' {\em Microwave Theory and Techniques, IEEE
  Transactions on}~{\bf 51}(10),  2139--2146 (2003).

\bibitem{potter}
Potter, P.,  [{\em A new horn antenna with suppressed sidelobes and equal
  bandwidths}{\nolinebreak\hspace{0.1em}]}, Jet Propulsion Laboratory,
  California Institute of Technology (1963).

\bibitem{zeng}
Zeng, L., Bennett, C.~L., Chuss, D.~T., and Wollack, E.~J., ``A low
  cross-polarization smooth-walled horn with improved bandwidth,'' {\em
  Antennas and Propagation, IEEE Transactions on}~{\bf 58}(4),  1383--1387
  (2010).

\bibitem{zhang2009}
Zhang, J., Ade, P., Mauskopf, P., Moncelsi, L., Savini, G., and Whitehouse, N.,
  ``New artificial dielectric metamaterial and its application as a terahertz
  antireflection coating,'' {\em Applied optics}~{\bf 48}(35),  6635--6642
  (2009).

\end{thebibliography}
\bibliographystyle{spiebib}   

\end{document}